\documentclass[10pt, conference, compsocconf]{IEEEtran}

\usepackage{graphicx}
\usepackage{epsfig}
\usepackage{amssymb}
\usepackage{amsfonts}
\usepackage{setspace}
\usepackage{subfigure}
\usepackage{algorithm2e}
\usepackage{amsmath}

\def\eq#1{\begin{equation} #1 \end{equation}}
\newcommand{\vect}[1]{\boldsymbol{#1}}
\def\xym{\{x_i,y_i\}}
\newcommand{\parsv}{\boldsymbol{\theta}}

\def\astroML{{\tt astroML}}

\begin{document}
%
\title{Introduction to {\tt astroML}: Machine Learning for Astrophysics}


\author{
\IEEEauthorblockN{Jacob VanderPlas}
\IEEEauthorblockA{Department of Astronomy\\
University of Washington \\
Seattle, WA 98155, USA\\
vanderplas@astro.washington.edu}
\and
\IEEEauthorblockN{Andrew J. Connolly}
\IEEEauthorblockA{Department of Astronomy\\
University of Washington \\
Seattle, WA 98155, USA\\
ajc@astro...}
\and
\IEEEauthorblockN{\v{Z}eljko Ivezi\'{c}}
\IEEEauthorblockA{Department of Astronomy\\
University of Washington \\
Seattle, WA 98155, USA\\
ivezic@astro...}
\and
\IEEEauthorblockN{Alex Gray}
\IEEEauthorblockA{College of Computing\\
Georgia Institute of Technology \\
Atlanta, GA 30332, USA \\
agray@cc.gatech.edu}
}


%

\maketitle

\begin{abstract}
  Astronomy and astrophysics are witnessing dramatic increases in data
  volume as detectors, telescopes and computers become ever more
  powerful. During the last decade, sky surveys across the
  electromagnetic spectrum have collected hundreds of terabytes of
  astronomical data for hundreds of millions of sources. Over the next
  decade, the data volume will enter the petabyte domain, and provide
  accurate measurements for billions of sources. Astronomy and 
  physics students are not traditionally trained to handle such voluminous 
  and complex data sets. In this paper we
  describe \astroML; an initiative, based on {\tt python} and {\tt scikit-learn},
  to develop a compendium of machine learning tools designed to
  address the statistical needs of the next generation of students and
  astronomical surveys. We introduce \astroML\ and present a number of 
  example applications that are enabled by this package.
\end{abstract}

\section{Introduction}

{\it Data mining}, {\it machine learning} and {\it knowledge
  discovery} are fields related to statistics, and to each
other. Their common themes are analysis and interpretation of data,
often involving large quantities of data, and even more often
resorting to numerical methods. The rapid development of these fields
over the last few decades was led by computer scientists, often in
collaboration with statisticians, and is built upon firm statistical
foundations. To an outsider, data mining, machine learning and
knowledge discovery compared to statistics are akin to engineering
compared to fundamental physics and chemistry: applied fields that
``make things work''.

Here we introduce \astroML, a {\tt python} package developed for extracting
knowledge from data, where ``knowledge'' means a quantitative summary of
data behavior, and ``data'' essentially means results of measurements.
Rather than re-implementing common data processing techniques,
\astroML\ leverages the powerful tools available in
{\tt numpy}\footnote{http://numpy.scipy.org},
{\tt scipy}\footnote{http://www.scipy.org},
{\tt matplotlib}\footnote{http://matplotlib.sourceforge.net},
{\tt scikit-learn}\footnote{http://scikit-learn.org} \cite{scikit-learn},
and other open-source packages, adding implementations of algorithms
specific to astronomy.
Its purpose is twofold: to provide an open repository for fast
{\tt python} implementations of statistical routines commonly used in
astronomy, and to provide accessible examples of astrophysical data
analysis using techniques developed in the fields of statistics and
machine learning.

The \astroML\ package is publicly
available\footnote{See http://ssg.astro.washington.edu/astroML/} and includes
dataset loaders, statistical tools and hundreds of example scripts.
In the following sections we provide a few  examples of how 
machine learning can be applied to astrophysical data using
\astroML\footnote{\astroML\ was designed to support a textbook
  entitled ``Statistics, Data Mining and Machine Learning in Astronomy''
  by the authors of this paper, to be published in 2013 by 
  Princeton University Press; these examples are adapted from this
  book.}.
We focus on regression (\S\ref{sec:regression}),
density estimation (\S\ref{sec:density}),
dimensionality reduction (\S\ref{sec:dimensionality}),
periodic time series analysis (\S\ref{sec:timeseries}),
and hierarchical clustering (\S\ref{sec:clustering}).
All examples herein are implemented using algorithms and datasets available
in the \astroML\ software package.

\section{Regression and Model Fitting}
\label{sec:regression}

Regression is a special case of the general model fitting problem.
It can be defined as the relation between a dependent
variable, $y$, and a set of independent variables, $x$, that describes
the expectation value of $y$ given $x$: $E[y|x]$. The solution to this
generalized problem of regression is, however, quite
elusive. Techniques used in regression therefore tend to make a
number of simplifying assumptions about the nature of the data, the
uncertainties on the measurements, and the complexity of the
models. An example of this is the use of regularization, discussed below.

The posterior pdf for the regression can be written as, 
\eq{ p(\vect{\theta}|\xym,I) \propto p(\xym | \vect{\theta}, I) \,
  p(\vect{\theta}, I).  }
Here the information $I$ describes the error behavior for the dependent
variable, and $\vect{\theta}$ are model parameters. The data likelihood 
is the product of likelihoods for individual points, and the latter can be 
expressed as 
\eq{
\label{eq:regressionBayesDL}
p(\{x_i,y_i\}|\vect{\theta}, I) = e(y_i|y) } 
where $y=f(x|\vect{\theta})$ is the adopted model class, and
$e(y_i|y)$ is the probability of observing $y_i$ given the true value
(or the model prediction) of $y$. For example, if the $y$
error distribution is Gaussian, with the width for $i$-th data point
given by $\sigma_i$, and the errors on $x$ are negligible, then 
\eq{
\label{eq:regressionError}
e(y_i|y) = {1 \over \sigma_i \sqrt{2\pi}} \,
\exp{\left({-[y_i-f(x_i|\vect{\theta})]^2 \over 2 \sigma_i^2}\right)}.
} 

It can be shown that under most circumstances
the least-squares approach to regression
results in the best unbiased estimator for the linear
model. In some cases, however, the regression problem may be ill-posed
and the best unbiased estimator is not the most appropriate regression
(we trade an increase in bias for a reduction in variance). Examples
of this include cases where attributes within high-dimensional
data show strong correlations
within the parameter space
(which can result in ill-conditioned matrices),
or when the number of terms in the regression model reduces the number of
degrees of freedom such that we must worry about overfitting of the
data.

One solution to these problems is to constrain or limit the complexity
of the underlying regression model. In a Bayesian framework, this can be
accomplished through the use of a non-uniform prior.
In a frequentist framework,
this is often referred to as regularization, or shrinkage, and works
by applying a penalty to the likelihood function.
Regularization can come in many forms, but is
usually a constraint on the smoothness of the model, or a limit on the
quantity or magnitude of the regression coefficients. We can impose
a penalty on this minimization if we include a regularization term,
\begin{equation}
  \label{eq:regress:penal}
  \chi^2(\parsv) = (Y - \parsv M)^T(Y- \parsv M) - \lambda |\parsv|^2,
\end{equation}
where $M$ is the design matrix that describes the regression model, 
$\lambda$ is a Lagrange multiplier, and $|\parsv |^2$ is an example the
penalty function. In this example, we penalize the size of the
regression coefficients. Minimizing $\chi^2$ w.r.t. $\parsv$ gives
\begin{equation}
  \parsv = (M^T C^{-1} M + \lambda I)^{-1} (M^T C^{-1} Y)
\end{equation}
where $I$ is the identity matrix and $C$ is a covariance matrix whose diagonal 
elements contain the uncertainties on the dependent variable, $Y$.

This regularization is often referred to as ridge regression or
Tikhonov regularization \cite{Tikhonov1995}. It provides a constraint
on the sum of the squares of the model coefficients, such that
\eq{ |\parsv |^2 < s, } where $s$ controls the complexity of the model
in the same way as the Lagrange multiplier $\lambda$ in
eqn.~\ref{eq:regress:penal}. By suppressing large regression
coefficients this constraint limits the variance of the system at the
expense of an increase in the bias of the derived coefficients.

\begin{figure}
  \includegraphics[width=0.5\textwidth]{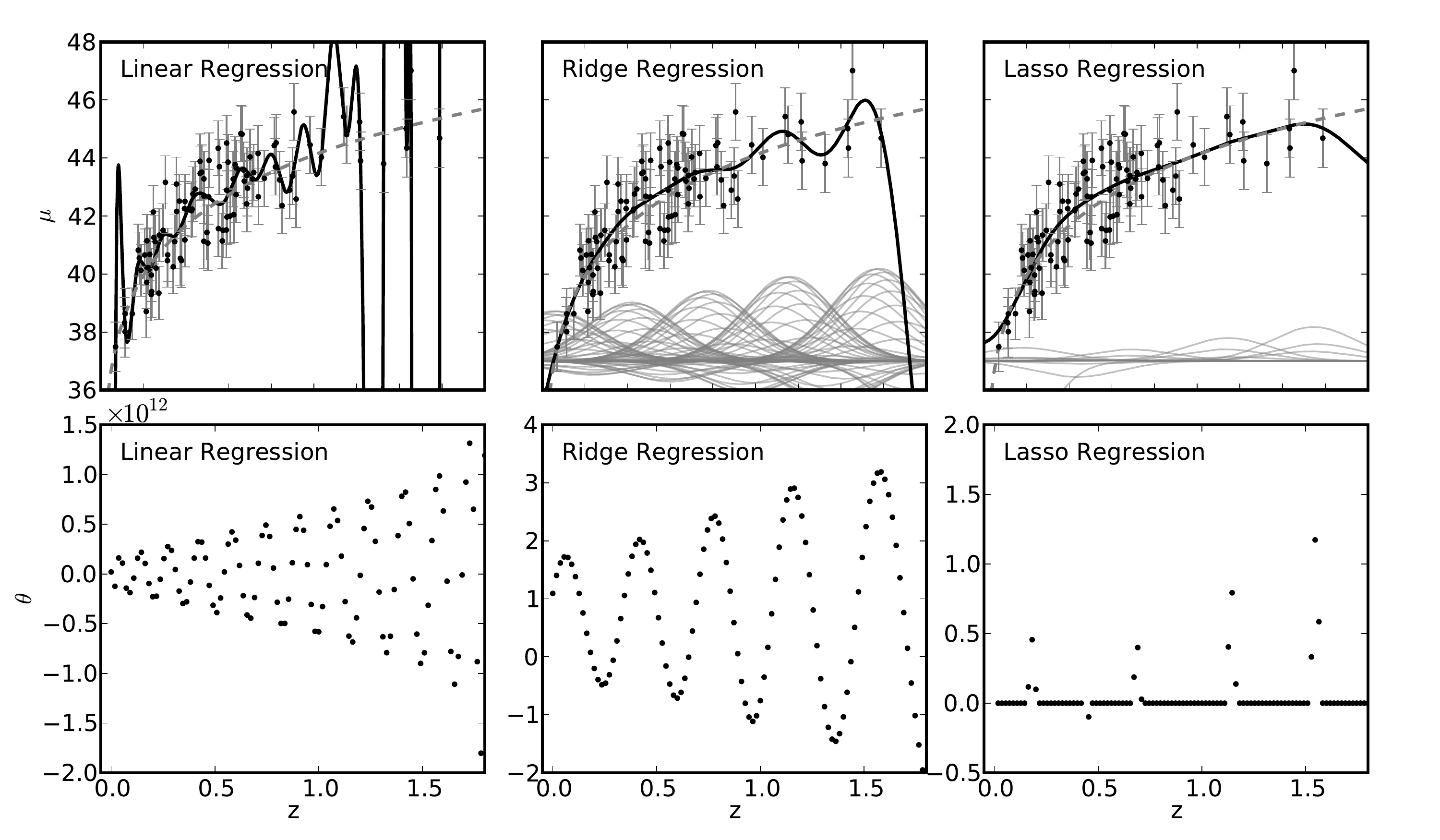}
  \caption{Example of a regularized regression for a simulated supernova
    dataset $\mu(z)$. 
    We use Gaussian Basis Function regression
    with a Gaussian of width $\sigma = 0.2$ centered at 100 regular intervals
    between $0 \le z \le 2$.  The lower panels show the best-fit weights as
    a function of basis function position.
    The left column shows the results with no regularization: the basis function
    weights $w$ are on the order of $10^8$, and over-fitting is evident.
    The middle column shows Ridge regression ($L_2$ regularization) with
    $\alpha = 0.005$, and the right column shows Lasso regression
    ($L_1$ regularization) with $\alpha = 0.005$.  All three fits include
    the bias term (intercept).  Dashed lines show the input curve.}
  \label{fig:mu_z_rbf_ridge}
\end{figure}

Figure~\ref{fig:mu_z_rbf_ridge} uses Gaussian basis function
regression  to illustrate how ridge regression constrains the
regression coefficients for simulated supernova data.  We fit $\mu$, a measure
of the distance based on brightness, vs $z$, a measure of the distance based
on the expansion of space.  The form of this relationship can yield insight
into the geometry and dynamics of the expanding universe:
similar observations
led to the discovery of dark energy \cite{Riess1998, Perlmutter1999}.
The left panel of Figure~\ref{fig:mu_z_rbf_ridge} shows a general linear
regression using 100 evenly spaced Gaussians as basis functions. This
large number of model parameters results in an overfitting
of the data, which is particularly evident at either end of the
interval where data is sparsely sampled. This overfitting is reflected
in the lower panel of Figure~\ref{fig:mu_z_rbf_ridge}: the
regression coefficients for this fit are on the order of 10$^8$.
The central panel demonstrates how ridge regression (with $\lambda
=0.005$) suppresses the amplitudes of the regression coefficients and
the resulting fluctuations in the modeled response.

A modification of ridge regression is to use the $L_1$-norm 
to impose sparsity in the model as well as apply shrinkage.
This technique is known as Lasso (least absolute shrinkage and
selection \cite{Tibshirani1996Regression}). Lasso penalizes the
likelihood as,
\begin{equation}
  \chi^2(\parsv) = (Y - \parsv M)^T(Y- \parsv M) - \lambda |\parsv|,
\end{equation} 
where $|\parsv|$ constrains the absolute value of $\parsv$. Lasso
regularization is equivalent to least squares regression within a
constraint on the absolute value of the regression coefficients \eq{ |
  \parsv| < s. }

The most interesting aspect of Lasso is that it not only weights the
regression coefficients, it also imposes sparsity on the regression
model.  This corresponds to
setting one (or more if we are working in higher dimensions) of the
model attributes to zero.  This subsetting of the model attributes
reduces the underlying complexity of the model (i.e.\ we force the model
to select a smaller number of features through zeroing of weights).
As $\lambda$ increases, the size of the region of parameter space
encompassed within the constraint decreases.
Figure~\ref{fig:mu_z_rbf_ridge}  illustrates this effect for our simulated
dataset: of the 100 Gaussians in the input model, with $\lambda=0.005$,
only 14 are selected by Lasso (note the regression coefficients in the
lower panel). This reduction in model complexity suppresses the
overfitting of the data.

In practice, regression applications in astronomy are rarely clean and
straight-forward.  Heteroscedastic measurement errors and missing or censored
data can cause problems.  Additionally, the form of the underlying regression
model must be carefully chosen such that it can accurately reflect the
fundamental nature of the data.

\section{Density Estimation Using Gaussian Mixtures} 
\label{sec:density}

The inference of the probability density distribution function (pdf) from a sample of data is 
known as density estimation. Density estimation is one of the most critical components of
extracting knowledge from data. For example, given a single pdf we can generate
simulated distributions of data and  compare against observations.
If we can identify regions of low probability within the pdf we have
a mechanism for the detection of unusual or anomalous sources. 

A common pdf model is the {\it Gaussian mixture model}, which describes a pdf
by a sum of (often multivariate) Gaussians. The optimization of the model
likelihood is typically done using the iterative
Expectation-Maximization algorithm \cite{DLR77}. 
Gaussian mixtures in the presence of data errors are known in astronomy as
Extreme Deconvolution (XD) \cite{bovy2011extreme}.  XD generalizes the EM approach 
to a case with measurement errors and possible pre-projection of the underlying
data. More explicitly, one assumes that  the noisy observations
$\mathbf{x}_i$ and the true values $\mathbf{v}_i$ are related through
\begin{equation}
  \mathbf{x}_i = \mathbf{R}_i \mathbf{v}_i + \mathbf{\epsilon}_i
\end{equation}
where $\mathbf{R}_i$ is the so-called projection matrix, which may be
rank-deficient. The noise $\mathbf{\epsilon}_i$ is assumed to be drawn
from a Gaussian with zero mean and variance $\mathbf{S}_i$. Given the
matrices $\mathbf{R}_i$ and $\mathbf{S}_i$, the aim of XD
is to find the model parameters describing the underlying Gaussians and their weights
in a way that maximizes the likelihood of the observed data. The EM approach to
this problem results in an iterative procedure that converges to (at
least) a local maximum of the likelihood. Details of the use of XD, including
methods to avoid local maxima in the likelihood surface, can be found
in \cite{bovy2011extreme}.

\begin{figure}
  \centering
  \includegraphics[width=0.5\textwidth]{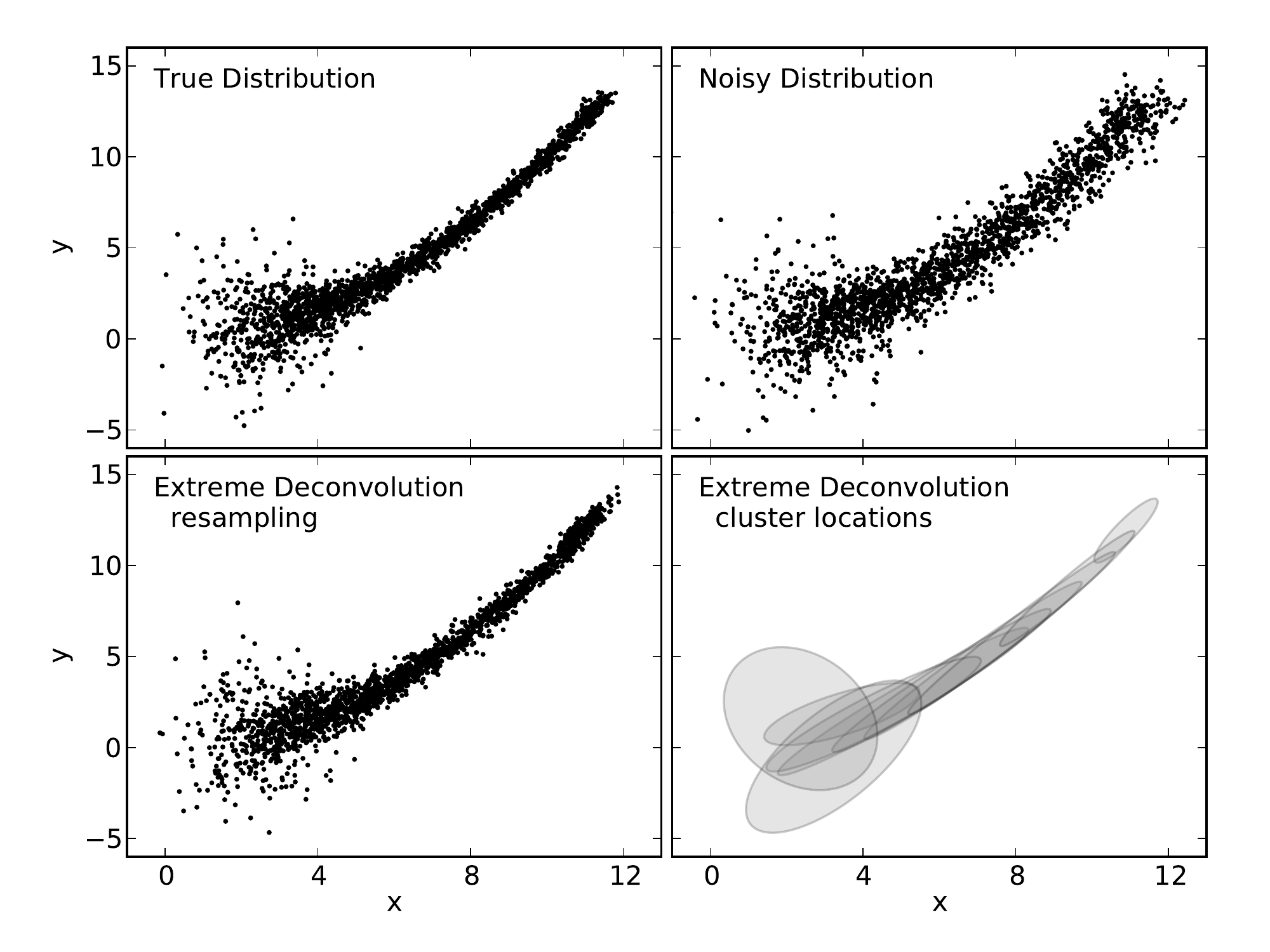}
  \caption{An example of extreme deconvolution showing the density of
    stars as a function of color from a simulated data set. The top
    two panels show the distributions for high
    signal-to-noise and lower signal-to-noise data. The lower panels
    show the densities derived from the noisy sample using
    extreme deconvolution; the resulting distribution closely matches
    that of the high signal-to-noise data.}
  \label{fig:XD_example}
\end{figure}

The XD implementation in \astroML\ is used for the example illustrated in
Figure~\ref{fig:XD_example}. The top panels show the true dataset (2000 points) and the 
dataset with noise added.  The bottom panels show the extreme deconvolution
results: on the left is a new dataset drawn from the mixture (as expected,
it has the same characteristics as the noiseless sample).  On the right
are the 2-$\sigma$ limits of the ten Gaussians used in the noisy data fit.
The important
feature of this figure is that from the noisy data, we are able to recover
a distribution that closely matches the true underlying data: we have
deconvolved the data and the noise.

In practice, one must be careful with XD that the measurement errors used in
the algorithm are accurate:  If they are over or under-estimated, the resulting
density estimate will not reflect that of the underlying distribution.
Additionally such EM approaches typically are only guaranteed to converge to
a {\it local} maximum.  Typically, several random initial configurations are
used to increase the probability of converging to a global maximum likelihood.

\section{Dimensionality of Data}
\label{sec:dimensionality}

Many astronomical analyses must address the question of the complexity
as well as size of the data set. For example, with imaging surveys
such as the LSST \cite{LSSToverview} and SDSS \cite{York2000}, we 
could measure arbitrary numbers of
properties or features for any source detected on an image (e.g.\ we
could measure a series of progressively higher moments of the
distribution of fluxes in the pixels that makeup the source). From the
perspective of efficiency we would clearly rather measure only those
properties that are directly correlated with the science we want to
achieve. In reality we do not know the correct measurement to use or
even the optimal set of functions or bases from which to construct
these measurements. Dimensionality reduction addresses these issues,
allowing one to search for the parameter combinations within a
multivariate data set that contain the most information.
 
\begin{figure}[htb]
\centering
\includegraphics[width=0.5\textwidth,angle=0]{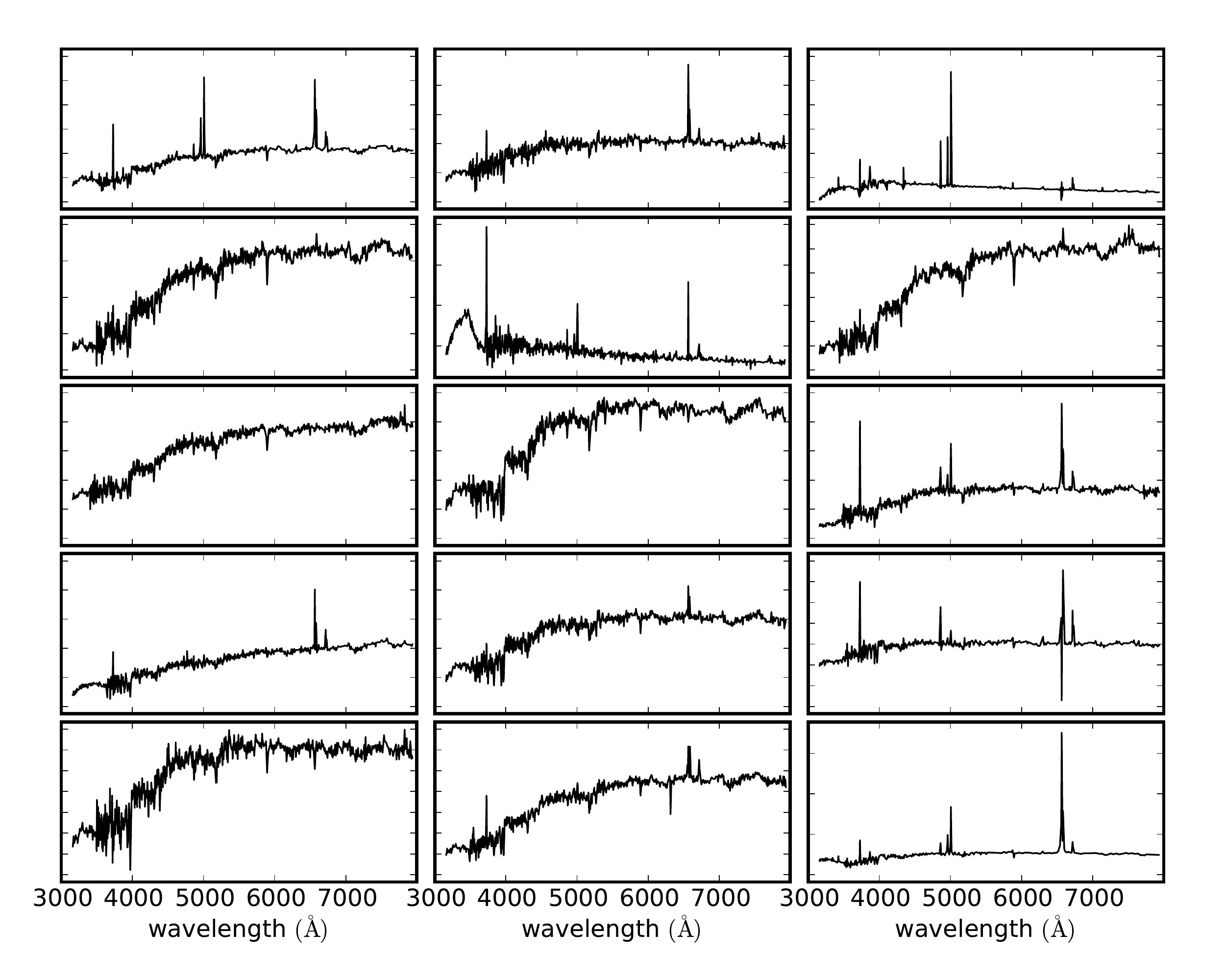}
\caption{A sample of fifteen galaxy spectra selected from the SDSS
  spectroscopic data set \cite{York2000}. These
  spectra span a range of galaxy types, from star-forming to passive
  galaxies. Each spectrum has been shifted to its rest-frame and covers
  the wavelength interval 3000--8000 \AA . The specific fluxes, $F_\lambda(\lambda)$,
  on the ordinate axes have an arbitrary scaling. }
\label{fig:sdss-spectra}
\end{figure}

We use SDSS galaxy spectra as an example of high dimensional data.
Figure~\ref{fig:sdss-spectra} shows a representative sample of these
spectra covering the rest-frame wavelength interval 3200--7800 \AA\ in 1000
bins. Classical approaches for identifying the principal dimensions
within the large samples of spectroscopic data include: principal
component analysis (PCA), independent component analysis (ICA), and 
non-negative matrix factorization (NMF). 

\subsection{Principal Component Analysis}
\begin{figure}\begin{center}
\includegraphics[width=0.5\textwidth,angle=0]{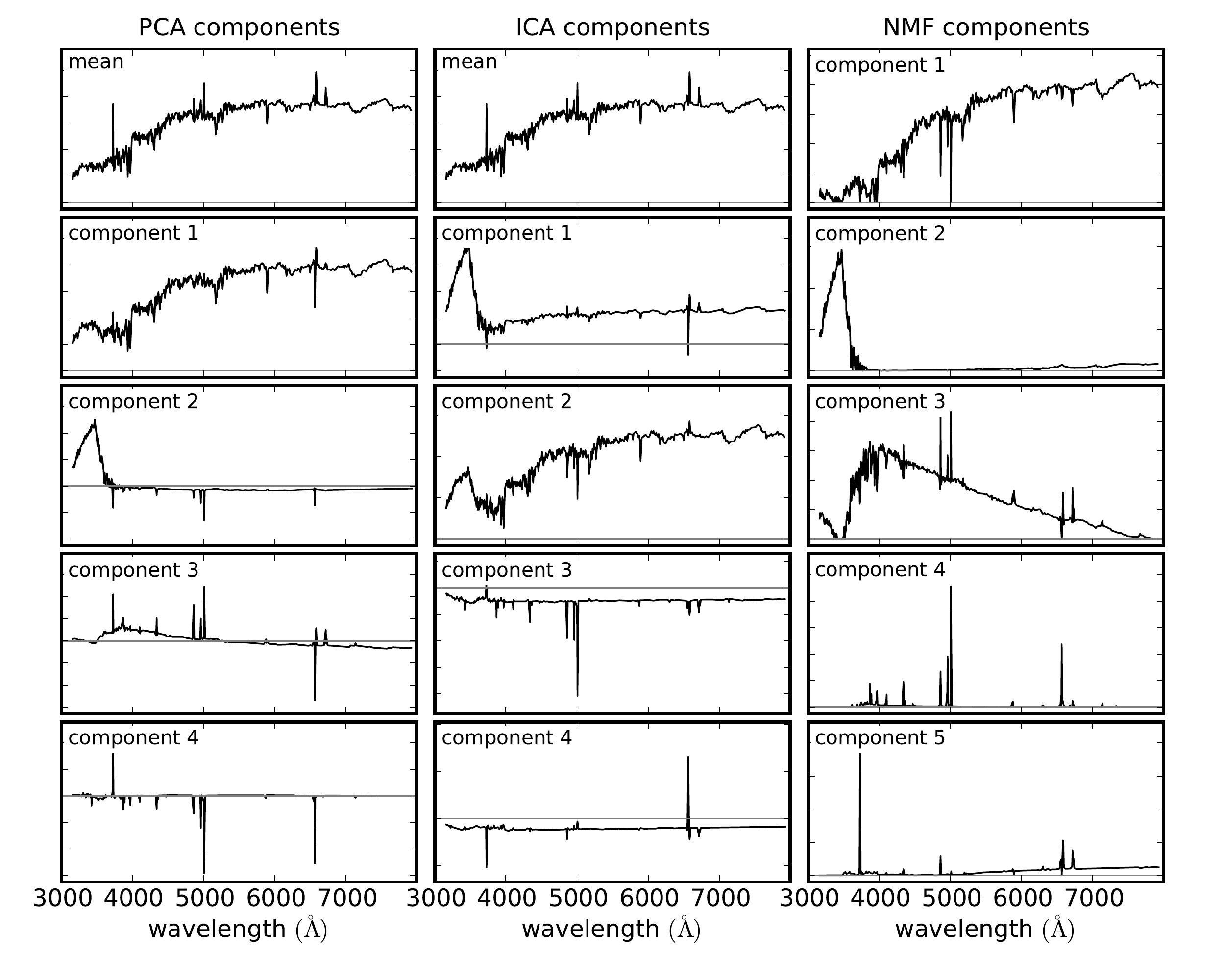}
\end{center}
\caption{A comparison of the decomposition of SDSS spectra using PCA
  (left panel), ICA (middle panel) and NMF (right panel). The rank
  of the component increases from top to bottom. For the ICA and PCA
  the first component is the mean spectrum (NMF does not require mean
  subtraction). All of these techniques isolate a common set of
  spectral features (identifying features associated with the
  continuum and line emission). The ordering of the spectral components is technique dependent.}
\label{fig:sdss-components}
\end{figure}

PCA is a linear transform, applied to multivariate data, that defines
a set of uncorrelated axes (the principal components) ordered by the
variance captured by each new axis.  It is one of the most widely
applied dimensionality reduction techniques used in astrophysics
today. Figure~\ref{fig:sdss-components} shows, from top to bottom, the
first four eigenvectors together with the mean spectrum.  The first 10
of the 1000 principal components represent 94\% of the total variance
of the system. From this we can infer that, with little loss of
information, we can represent each galaxy spectrum as a linear sum of
a small number of eigenvectors. Eigenvectors with large eigenvalues
are predominantly low order components (in the context of the
astronomical data they reflect the smooth {\it continuum} component of the
galaxies). Higher order components, which have smaller eigenvalues, are
predominantly made up of sharp features such as atomic emission lines,
or uncorrelated features like spectral noise.

As seen in Figure~\ref{fig:sdss-components}, the eigenspectra of SDSS galaxies
are reminiscent of the various stellar components which make up the galaxies.
However, one of the often-cited limitations of PCA is that the eigenvectors are
based on the variance, which does not necessarily reflect any true
physical properties of the data.  For this reason, it is useful to explore
other linear decomposition schemes.


\subsection{Independent Component Analysis}
Independent Component Analysis (ICA) \cite{Comon1994Independent} is a
computational technique that has become popular in the biomedical
signal processing community to solve what has often been referred to
as blind source-separation or the ``cocktail party problem''
\cite{cherry1953}. In this problem,
there are multiple microphones situated through out a room containing
$N$ people. Each microphone picks up a linear combination of the $N$
voices. The goal of ICA is to use the concept of statistical
independence to isolate (or unmix) the individual signals. 

The principal that underlies ICA comes from the observation that the
input signals, $s_i(\lambda)$, should be statistically
independent. Two random variables are considered statistically
independent if their joint probability distribution, $f(x,y)$, can be
fully described by a combination of their marginalized probabilities,
i.e.\,
\begin{equation}
  f(x^p,y^q)=f(x^p)f(y^q)
\end{equation}
where $p$, and $q$ represent arbitrary higher order moments of the
probability distributions \cite{stonePorrill1997}.

ICA refers to a class of related algorithms: in most of these the requirement
for statistical independence is expressed in terms of the non-Gaussianity
of the probability distributions. The rationale for this is that,
by the central limit theorem, the
sum of any two independent random variables will always be more
Gaussian than either of the individual random variables. This would
mean that, for the case of the independent stellar components that make up a
galaxy spectrum, the input signals can be approximated by identifying an
unmixing matrix, $W$, that maximizes the non-Gaussianity of the resulting
distributions. Definitions of non-Gaussianity include
the kurtosis of a distribution, the negentropy (the
negative of the entropy of a distribution), and mutual information.

The middle panel of Figure~\ref{fig:sdss-components} shows the ICA
components derived via the FastICA algorithm from the spectra represented in
Figure~\ref{fig:sdss-spectra}.
As with many multivariate applications, as the size of the mixing matrix
grows, computational complexity often makes it impractical to
calculate it directly. Reduction in the complexity of the input signals
through the use of PCA (either to filter the data or to project the
data onto these basis functions) is often applied to ICA applications.

\subsection{Non-negative Matrix Factorization}
Non-negative matrix factorization (NMF) is similar in spirit to PCA, 
but adds a positivity constraint on the components that comprise the
data matrix $X$ \cite{LeeDaniel2001Algorithms}. It assumes that any
data matrix can be factored into two matrices, $W$ and $Y$, such that,
\begin{equation}
  X=W Y,
\end{equation}
where both $W$ and $Y$ are non-negative (i.e.\ all elements in
these matrices are $\ge 0$). $WY$ is, therefore, an approximation of $X$. By
minimizing the reconstruction error $|| (X - W Y)^2 ||$
it is shown in
\cite{LeeDaniel2001Algorithms} that non-negative bases can be
derived using a simple update rule,
\begin{eqnarray}
\label{eqn:nmf}
W_{ki} &=& W_{ki}\frac{[XY^T]_{ki}}{[WYY^T]_{ki}},\\ 
Y_{in} &=& Y_{in}\frac{[W^TX]_{in}}{[W^TWY]_{in}},
\end{eqnarray}
where $n$, $k$, and $i$ denote the wavelength, spectrum and template
indices respectively.  This iterative process does not guarantee a
local minimum, but random initialization and cross-validation procedures can
be used to determine appropriate NMF bases.

The right panel of Figure~\ref{fig:sdss-components} shows the results of
NMF applied to the spectra shown in Figure~\ref{fig:sdss-spectra}.
Comparison of the components derived from PCA, ICA, and NMF in
Figure~\ref{fig:sdss-components} shows that these
decompositions each produce a set of basis functions that are broadly
similar to the others, including both continuum and line emission features.
The ordering of the importance of each component is dependent on the technique: 
in the case of ICA, finding a subset of ICA components is not the same
as finding all ICA components \cite{GirolamiFyfe1997}. The a priori
assumption of the number of underlying components will affect the form
of these components.

These types of dimensionality techniques can be useful for identifying classes
of objects, for detecting rare or outlying objects, and for constructing compact
representations of the distribution of observed objects.  One potential
weakness of dimensionality reduction algorithms is that the components are
defined statistically, and as such have no guarantee of reflecting true physical
aspects of the systems being observed.  Because of this, one must be
careful when making physical inferences from such results.

\section{Time Series Analysis}
\label{sec:timeseries}

Time series analysis is a branch of applied mathematics developed
mostly in the fields signal processing and statistics.  Even when limited
to astronomical datasets, the diversity of applications is
enormous. The most common problems range from detection of variability
and periodicity to treatment of non-periodic variability and searches
for localized events. The measurement errors can range from as small
as one part in 100,000, such as for photometry from the Kepler mission
\cite{Kepler}, to potential events buried in noise with a
signal-to-noise ratio of a few at best, such as in searches for
gravitational waves using the Laser Interferometric Gravitational
Observatory (LIGO) data \cite{LIGO}.  Datasets can include many
billions of data points, and sample sizes can be in millions, such as
for the LINEAR data set with 20 million light curves, each with a few
hundred measurements \cite{LINEAR}. The upcoming Gaia and LSST
surveys will increase existing datasets by large factors; the Gaia
satellite will measure about a billion sources close to 100 times
during its five year mission, and the ground-based LSST will obtain about
1000 measurements for about 20 billion sources during its ten years of
operations.  Scientific utilization of such datasets includes searches
for extrasolar planets, tests of stellar astrophysics through studies
of variable stars and supernovae explosions, distance determination,
and fundamental physics such as tests of general relativity with radio
pulsars, cosmological studies with supernovae and searches for
gravitational wave events.

One of the most popular tools for analysis of regularly (evenly)
sampled time series is the discrete Fourier transform. However, it
cannot be used when data are unevenly (irregularly) sampled. The
Lomb-Scargle periodogram \cite{Gottlieb1975, Lomb, Scargle} is a standard
method to search for periodicity in unevenly sampled time series data. 
The Lomb-Scargle periodogram corresponds to a single sinusoid model, 
\eq{
\label{eq:sineModel2}
            y(t) = a\, \sin(\omega t) + b \, \cos(\omega t),
} 
where $t$ is time and $\omega$ is angular frequency ($=2\pi f$). The model is linear with
respect to coefficients $a$ and $b$, and non-linear only with respect to frequency $\omega$. 
A Lomb-Scargle periodogram measures the power $P_{LS}(\omega)$,
which is a
straightforward trigonometric calculation involving the times,
amplitudes, and uncertainties of the observed quantity (details can be found
in, e.g.\ \cite{ZechKur}).  An important property of this technique is that the
periodogram $P_{LS}(\omega)$ is directly related to the $\chi^2$ of this
model evaluated with maximum a-posteriori estimates for $a$ and $b$
\cite{ZechKur, CMB1999}. It can be thought of as an ``inverted'' plot
of the $\chi^2(\omega)$ normalized by the ``no-variation'' $\chi^2_0$
($0 \le P_{LS}(\omega) < 1$).

There is an important practical deficiency in the original Lomb-Scargle method: 
it is implicitly assumed that the mean value of data values $\overline{y}$ is a 
good estimator of the mean of $y(t)$. In practice, the data often do not sample all 
the phases equally, the dataset may be small, or it may not extend over the whole
duration of a cycle; the resulting error in mean can cause problems such as aliasing
\cite{CMB1999}. A simple remedy proposed in \cite{CMB1999} is to add a constant
offset term to the model from eqn.~\ref{eq:sineModel2}.
Zechmeister and K\"{u}rster \cite{ZechKur} 
have derived an analytic treatment of this approach, dubbed the ``generalized'' Lomb-Scargle 
periodogram (it may be confusing that the same terminology was used by Bretthorst 
for a very different model \cite{Brett2001c}). The resulting expressions have a similar 
structure to the equations corresponding to standard Lomb-Scargle approach listed above 
and are not reproduced here.

\begin{figure}
  \centering
  \includegraphics[width=0.5\textwidth]{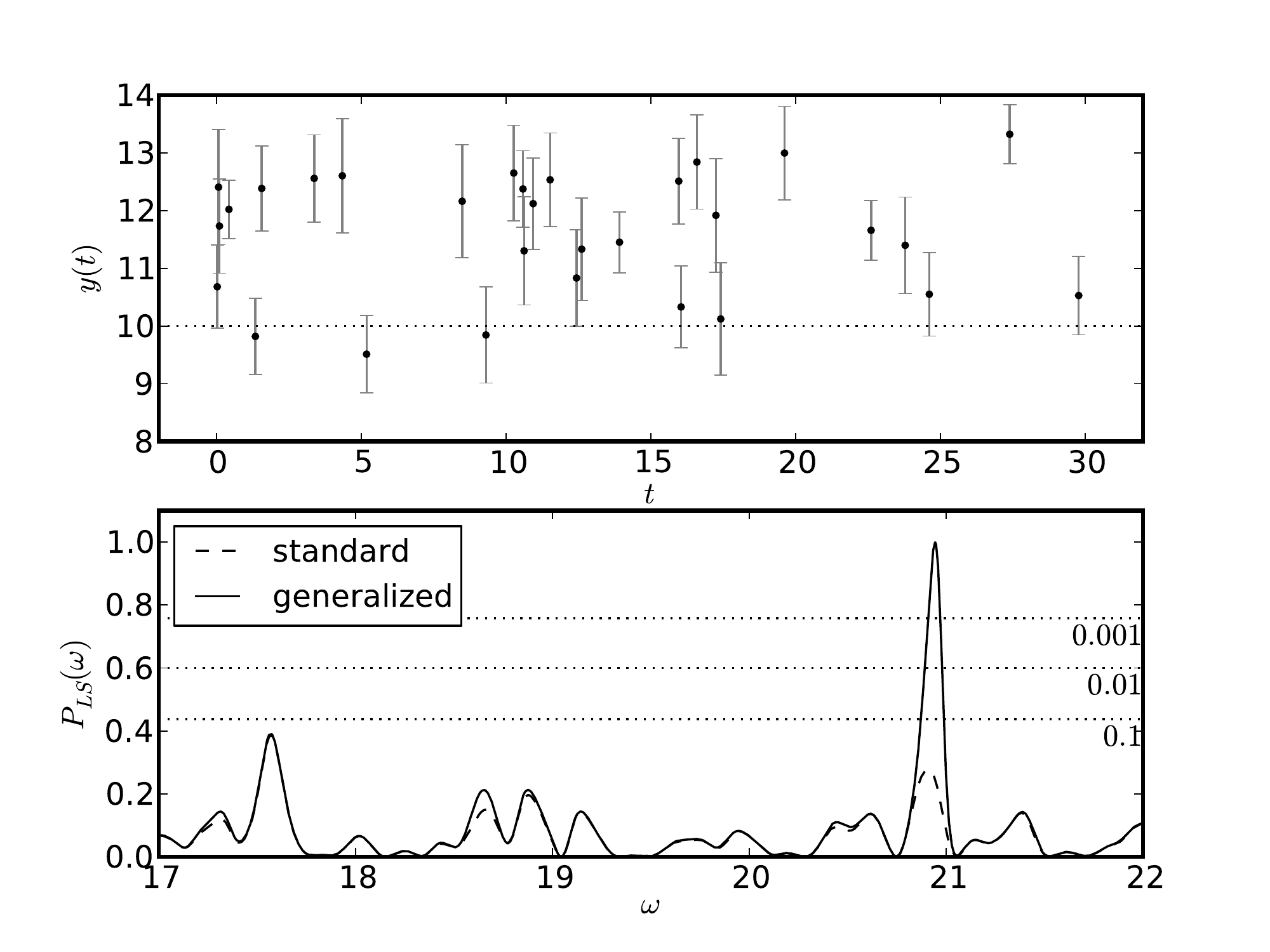}
  \caption{A comparison of standard and generalized Lomb-Scargle periodograms
    for a signal $y(t) = 10 + \sin(2\pi t/P)$ with $P=0.3$, corresponding
    to $\omega_0 \approx 21$.  This example is in some sense a worst-case
    scenario for the standard Lomb-Scargle algorithm because there
    are no sampled points during the times when $y_{true} < 10$, which
    leads to a gross overestimation of the mean.  The bottom panel shows the Lomb-Scargle
    and generalized Lomb-Scargle periodograms for these data: the generalized
    method recovers the expected peak, while the standard method misses the
    true peak and choose a spurious peak at $\omega \approx 17.6$.}
    \label{fig:standard_vs_general}
\end{figure}

Both the original and generalized Lomb-Scargle methods are implemented in 
\astroML. Figure~\ref{fig:standard_vs_general} compares the two in
a worst-case scenario where the data sampling is such that the standard
method grossly overestimates the mean.  While the standard approach fails to
detect the periodicity due to the unlucky data sampling, the generalized
Lomb-Scargle approach recovers the expected period of $\sim$0.3 days,
corresponding to $\omega \approx 21$.
Though this example is quite contrived, it is not entirely artificial:
in practice this situation could arise if the period
of the observed object were on the order of 1 day, such that minima
occur only in daylight hours during the period of observation.

The underlying model of the Lomb-Scargle periodogram is non-linear in
frequency and thus in practice the maximum of the periodogram is found
by grid search.  The searched frequency range can be bounded by
$\omega_{min} = 2 \pi / T_{data}$, where $T_{data}=t_{max} - t_{min}$ is the
interval sampled by the data, and by $\omega_{max}$. As a good choice for
the maximum search frequency, a pseudo-Nyquist frequency
$\omega_{max} = \pi \overline{1/\Delta t}$, where $\overline{1/\Delta t}$ 
is the median of the inverse time interval between data points,
was proposed by \cite{Debo}
(in case of even sampling,  $\omega_{max}$ is equal to the Nyquist frequency). 
In practice, this choice may be a gross underestimate because unevenly
sampled data can detect periodicity with frequencies even higher than
$2\pi / (\Delta t)_{min}$ \cite{EyerBartholdi}.  
An appropriate choice of $\omega_{max}$ thus depends on sampling (the phase 
coverage at a given frequency is the relevant quantity) and needs to be
carefully  chosen: a very conservative limit on maximum detectable frequency
is of course given by the time interval over which individual measurements
are performed, such as imaging exposure time. 

An additional pitfall of the Lomb-Scargle algorithm is that the
classic algorithm only fits a single harmonic to the data.  For more complicated
periodic data such as that of a double-eclipsing binary stellar system, this
single-component fit may lead to an alias of the true period.

\section{Hierarchical Clustering: Minimum Spanning Tree}
\label{sec:clustering}

{\it Clustering} is an approach to data analysis which seeks to discover
groups of similar points in parameter space.  Many clustering algorithms
have been developed; here we briefly explore a hierarchical clustering
model which finds clusters at all scales.
Hierarchical clustering can be approached as a {\em divisive} (top-down)
procedure, where the data is progressively sub-divided, or as an
{\em agglomerative} (bottom-up) procedure, where clusters are built by
progressively merging nearest pairs.  In the example below, we will use
the agglomerative approach.

We begin at the smallest scale with $N$ clusters, each consisting of a 
single point.  At each step in the clustering process we merge the
``nearest'' pair of clusters: this leaves $N-1$ clusters remaining.
This is repeated $N$ times so that a single cluster remains, encompassing
the entire data set.  Notice that if two points are in the same cluster
at level $m$, they remain together at all subsequent levels: this is the
sense in which the clustering is hierarchical.    This approach is similar to
``friends-of-friends'' approaches often used in the analysis of $N$-body
simulations \cite{Davis1985, Audit1998}.
A tree-based visualization of a hierarchical clustering model
is called a {\em dendrogram}.

At each step in the clustering process we merge the ``nearest''
pair of clusters.  Options for
defining the distance between two clusters, $C_k$ and $C_{k'}$, include:
\begin{eqnarray}
  d_{\rm {min}}(C_k,C_{k'}) & = & \min_{x \in C_k, x' \in C_{k'}} ||x-x'|| \\
  d_{\rm {max}}(C_k,C_{k'}) & = & \max_{x \in C_k, x' \in C_{k'}} ||x-x'|| \\
  d_{\rm {avg}}(C_k,C_{k'}) & = & \frac{1}{N_k N_{k'}} 
  \sum_{x \in C_k} \sum_{x' \in C_{k'}} ||x-x'|| \\
  d_{\rm {cen}}(C_k,C_{k'}) & = & ||\mu_k - \mu_{k'}||
\end{eqnarray}
where $x$ and $x'$ are the points in cluster $C_k$ and $C_{k'}$
respectively, $N_k$ and $N_{k'}$ are the number of points in each
cluster and $\mu_k$ the centroid of the clusters.

\begin{figure}
 \centering
 \includegraphics[width=0.5\textwidth]{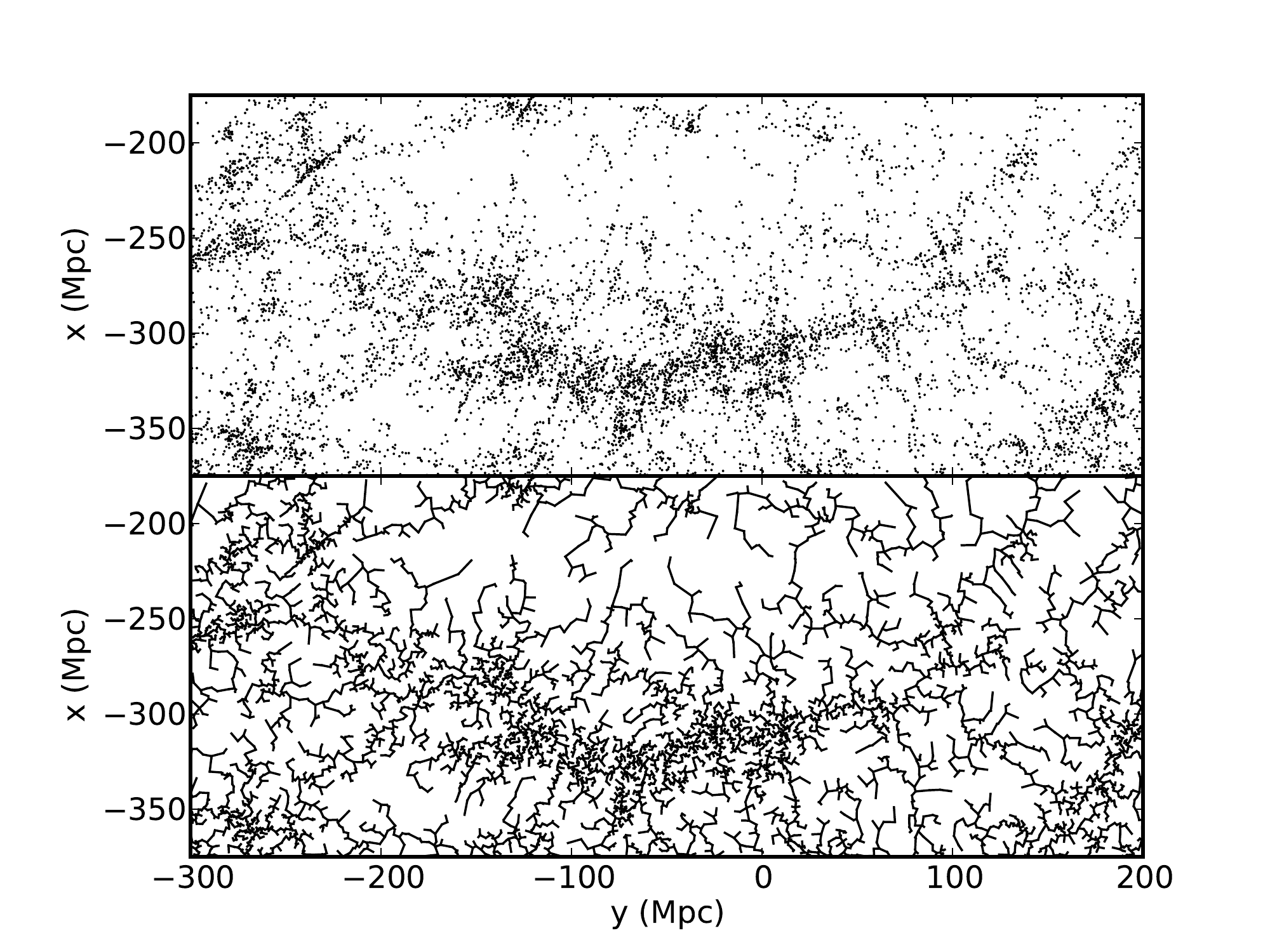}
 \caption{Dendrogram of a minimum spanning tree over the 2D projection
   of the SDSS Great Wall.  Each of the 8014 points represents the
   two-dimensional spatial location of a galaxy.}
 \label{fig:great_wall_mst}
\end{figure}

Using the distance $d_{\rm {min}}$ results in a hierarchical
clustering known as a minimum spanning tree (see
\cite{Barrow1985, Krzewina1996, Allison2009}, for some
astronomical applications) and will
commonly produce clusters with extended chains of points. Using
$d_{\rm {max}}$ tends to produce hierarchical clustering with compact
clusters. The other two distance examples have behavior somewhere
between these two extremes.

Figure~\ref{fig:great_wall_mst} shows a dendrogram for a
hierarchical clustering model using a minimum spanning tree.
The data is the SDSS ``Great Wall'', a filament of galaxies that
is over 100 Mpc in extent \cite{Gott}.
The extended chains of points trace the large scale structure present
within the data. Individual clusters can be isolated by sorting the edges
by increasing length, then removing edges longer than some threshold.
The clusters consist of the remaining connected groups.

Unfortunately a minimum spanning tree is $O(N^3)$ to compute using
a brute-force approach, though this can be improved using well-known
graph traversal algorithms.
\astroML\ implements a fast approximate minimum spanning tree based on
sparse graph representation of links among nearest neighbors, using
{\tt scipy}'s sparse graph submodule.

\section{Discussion}
\label{sec:discussion}
The above examples are just a small subset of the analysis methods covered
in the \astroML\ codebase.  Each of these methods has advantages and
disadvantages in practice.  A full discussion of these characteristics
is beyond the scope of this work; we suggest a careful reading of the
references both in this paper and in the \astroML\ code.

To some extent, \astroML, and the upcoming companion book, are analogous
to the well-known Numerical Recipes book, but aimed at the analysis of massive 
astronomical data sets with emphasis on modern tools for data mining and
machine learning.  A strength of \astroML\ lies in the fact that
the {\tt python} code used to download, process, analyze, and plot the data
is all open-source and freely available, and that the examples are geared
toward common applications in the field of astronomy.

Rather than creating a heavy-weight comprehensive package which duplicates
many tools already available in well-known open-source {\tt python} libraries,
\astroML\ places a priority on maintaining a light-weight codebase and
using existing tools and packages when available.  Notably, code developed
for use in \astroML\ has already been submitted
upstream, and is included in the latest releases of {\tt scipy} and
{\tt scikit-learn}, where it is being used by researchers in fields
beyond astronomy and astrophysics (notable examples are the {\tt BallTree} in
{\tt scikit-learn}, and the sparse graph module in {\tt scipy}).
This open-source model of code sharing and development is an increasingly
important component of reproducibility of results in the age of data-driven
science.  We invite all colleagues who are currently developing software for
analysis of massive survey data to contribute to the \astroML\ code base.
With this sort of participation, we hope for \astroML\ to become a
community-driven resource for research and education in data-intensive
science.





\bibliographystyle{IEEEtran}
\bibliography{astroml}

\end{document}